\documentclass[12pt]{article}
\usepackage{epsf}
\usepackage{graphicx}
\usepackage{psfig,color,epsfig}
\usepackage{eepic}
\usepackage{latexsym}
\usepackage{amsmath}
\newcommand{\be}{\begin{equation}}
\newcommand{\ee}{\end{equation}}
\newcommand{\ba}{\begin{array}}
\newcommand{\ea}{\end{array}}
\newcommand{\bqa}{\begin{eqnarray}}
\newcommand{\eqa}{\end{eqnarray}}
\begin{document}
\begin{center}
{\Large\bf  The pole structure of the unitary, crossing symmetric
low energy $\pi\pi$ scattering amplitudes  }
\\[10mm]
{\sc Z.~Y.~Zhou,  G.~Y.~Qin\footnote{Present address: Department
of Physics, McGill University, 3600 University Street, Montreal,
Quebec, H3A 2T8, Canada. }, P.~Zhang,  Z.~G.~Xiao and H.~Q.~Zheng}
\\[2mm]
{\it  Department of Physics, Peking University, Beijing 100871,
P.~R.~China }
\\[2mm] and \\[2mm]
{\sc N.~Wu}
\\[2mm]
{\it  Institute for High Energy Physics, Chinese Academy of
Science, Beijing 100039, P.~R.~China }
\\[5mm]
\today
\begin{abstract}
The pole structure of the low energy $\pi\pi$ scattering
amplitudes is studied  using a proper chiral unitarization method
combined with crossing symmetry and the low energy phase shift
data. It is found that the $\sigma$ pole position is  at
$M_\sigma=470\pm 50MeV$, $\Gamma_\sigma=570\pm 50MeV$. The
existence of the virtual state pole in the IJ=20 channel is
reconfirmed. Various threshold parameters are estimated and are
found in general in good agreement with the results obtained from
the Roy equation analyses.
\end{abstract}
\end{center}
\vspace{1cm}

The $\sigma$  or the $f_0(600)$ resonance has attracted much
interest in recent years~\cite{review} and more and more
physicists, if not all, believe that it truly exists as a light
and broad resonance.~\cite{PDG02} There were two main difficulties
to accept the $\sigma$ resonance, one is that being light and
broad makes the $\sigma$'s contribution to the phase shift hard to
be distinguishable from background contributions while the latter
are often out of control in various phenomenological studies.
However, a careful analysis revealed that the cut contribution to
the phase shift is negative in the I,J=0,0 channel $\pi\pi$
scattering~\cite{XZ00}, and thus the $\sigma$ resonance has to
exist to saturate the experimental data.
 Another
objection is that the $\pi\pi$ phase shift in the I=0, J=0 channel
does not pass $\pi/2$ at moderately low energies which should have
appeared if there exist a light $\sigma$ as required by the
standard Breit--Wigner description of resonances. However it is
pointed out that a light and broad resonance rejects a simple
Breit-Wigner description. As a consequence a light and broad
resonance's contribution to the phase shift may pass 90$^\circ$
only at energies well above 1GeV.~\cite{Ztalk03}

Although the existence of the $\sigma$ resonance seems to be
firmly established, a precise determination to the location of
this resonance on the complex $s$ plane can still be challenging.
A summary of the previous determinations on the pole location of
the $\sigma$ resonance may be found in Ref.~\cite{ML99} and the
results are found to be rather diverse, which is reflected in the
Review of Particle Properties as the mass of the $\sigma$ ranges
from 400MeV to 1200MeV. In more recent studies, the pole mass of
the $\sigma$ resonance is estimated to be $M\sim 500$MeV and
$\Gamma\sim 640$MeV in Ref.~\cite{XZ00}, in Ref.~\cite{CGL01} an
extensive analysis using Roy equations indicates $M=470\pm 30$MeV,
$\Gamma=590\pm 40$MeV. On the other side, the E791 experimental
group claim from the $D^+\to\pi^+\pi^-\pi^+$ data analysis that
$M=478^{+24}_{-23}\pm 17$MeV, $\Gamma=324^{+42}_{-40}\pm
21$MeV,\footnote{Corresponding to a pole mass: $M=489$MeV,
$\Gamma=346$MeV. }~\cite{E791} whereas the BES collaboration has
found from BES-II $J/\psi\to \omega\pi\pi$ data the $\sigma$ pole
mass to be $M=541\pm 20\pm 32$MeV, $\Gamma=504\pm 50\pm
60$MeV~\cite{BESII} and BES-I $J/\psi\to \omega\pi\pi$ data the
pole mass as $M=434\pm 78$MeV, $\Gamma=404\pm 86$MeV.~\cite{BESI}
Under this situation it is worthwhile to make further efforts
towards a more accurate determination to the $\sigma$ pole. What
we will adopt here is a new approach recently proposed by our
group which is elaborated in details in Ref.~\cite{piK} discussing
$\pi K$ scatterings. The approach respects all known fundamental
properties of  $S$ matrix theory except that crossing symmetry is
not implemented automatically. We will remedy the deficiency here
by considering the constraints from crossing symmetry, the so
called Balanchandran-Nuyts-Roskies (BNR) relations in the fit. In
the following we begin by  briefly review the new unitarization
scheme developed in Ref.~\cite{piK}.

\begin{figure}
\mbox{\epsfxsize=6.5cm\epsffile{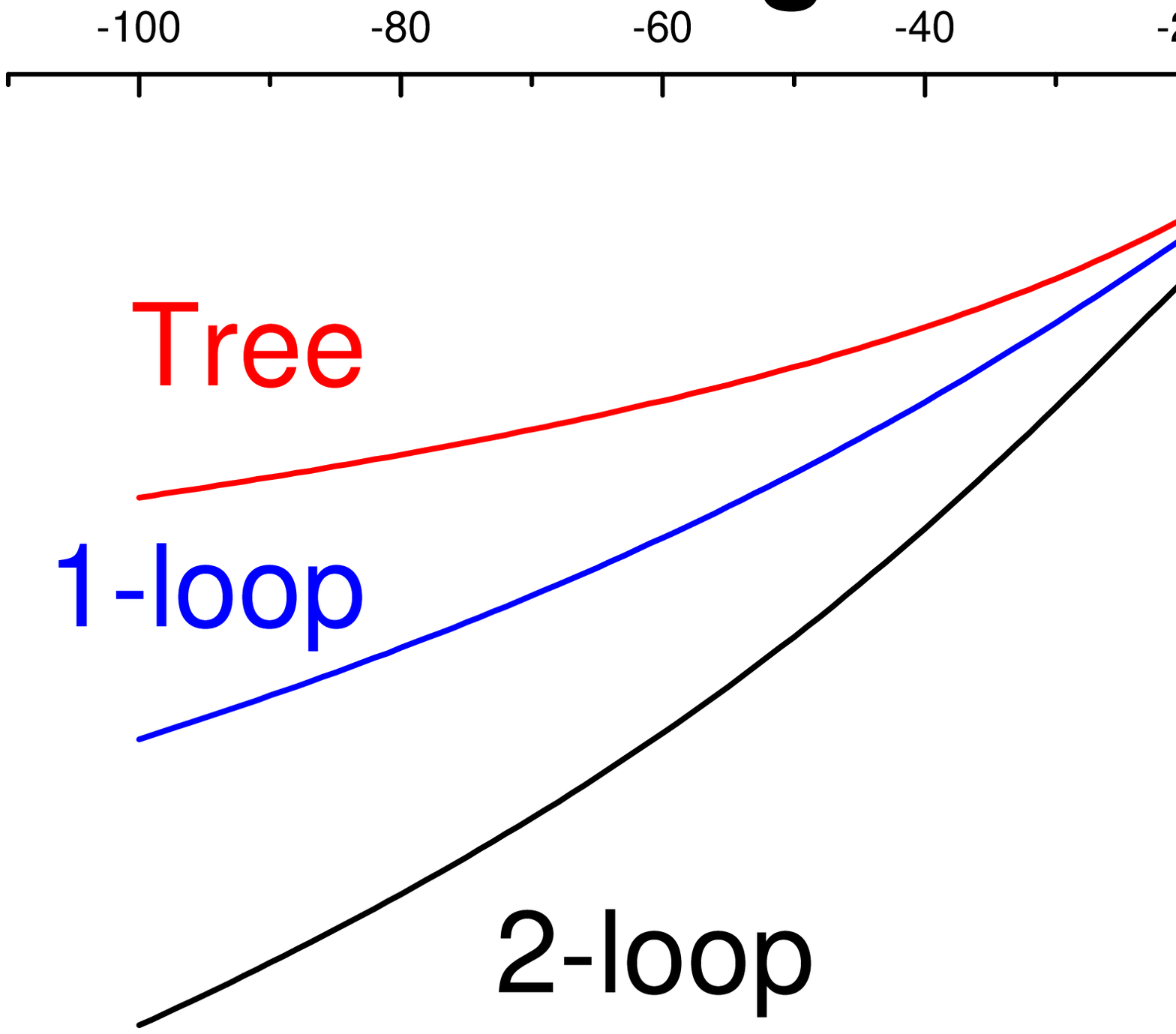}\hspace{1cm}\epsfxsize=6.5cm\epsffile{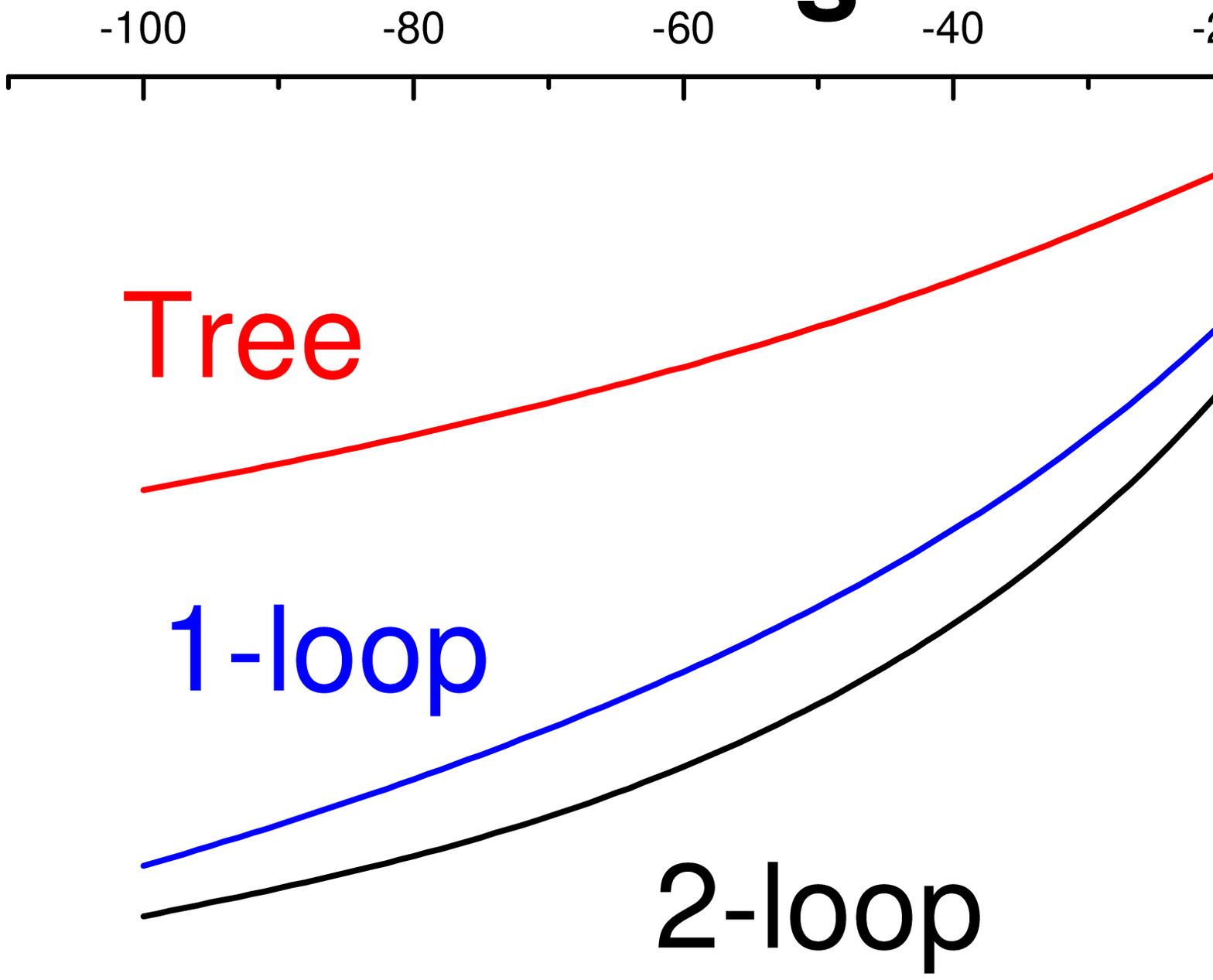}}
\vspace{-2cm}\caption{\label{ImLf}Tree, 1--loop and 2--loop
results of $\mathrm{Im}_{L}f(s)$ from Eq.~(\ref{IMLR'}): left)
I=0, J=0; right) I=2, J=0. The low energy parameters of the
$O(p^6)$ chiral Lagrangian are taken from Ref.~\cite{2loops}. }
\end{figure}

For a partial wave scattering in a given channel, the physical $S$
matrix can in general be factorized as,
 \be\label{param}
S^{phy}=\prod_iS^{p_i}\cdot S^{cut}\ ,
 \ee
 where $S^{p_i}$ are
the simplest $S$ matrices characterizing the isolated
singularities of $S^{phy}$~\cite{piK}, which are:
\begin{enumerate}
\item for a virtual state pole at $s=s_0$ ($0<s_0<4m_\pi^2$),
  \be \label{vsp} S^v(s)=\frac{1+i\rho(s)a }{1-i\rho(s)a}\
. \ee The scattering length contributed by a  virtual state pole
defined as such is
 \be\label{v_s_l}
  a^{v}=\sqrt{s_0\over 4m_\pi^2-s_0}\ .
  \ee
 \item for a  resonance located at $z_0$ (and
$z_0^*$) on the second sheet, the $S$ matrix can be written as,
  {\be\label{resp}
S^R(s)=\frac{M^2[z_0]-s+i\rho(s)s\frac{{\rm Im}[z_0]}{{\rm
Re}[\sqrt{z_0(z_0-4m_\pi^2)}]}}{M^2[z_0]-s-i\rho(s)s\frac{{\rm
Im}[z_0]}{{\rm Re}[\sqrt{z_0(z_0-4m_\pi^2)}]}}\ , \ee } where {
\be M^2[z_0]={\rm Re}[z_0] + {\rm Im}[z_0]\frac{{\rm
Im}[\sqrt{z_0(z_0-4m_\pi^2)}]}{{\rm
Re}[\sqrt{z_0(z_0-4m_\pi^2)}]}\ .\ee}
\end{enumerate}
The expression of the scattering length contributed by a resonance
defined as above is also easily obtainable.  The $S^{cut}$
contains only cuts which can be parameterized in the following
simple form, \bqa\label{fS'}
 S^{cut}&=&e^{2i\rho f(s)}\ ,\\
 f(s)&=&f(s_0)+\frac{s-s_0}{\pi}\int_{L}\frac{{\rm
 Im}_Lf(s')}{(s'-s_0)(s'-s)}ds'\nonumber\\
 &&+\frac{s-s_0}{\pi}\int_{R}\frac{{\rm
 Im}_Rf(s')}{(s'-s_0)(s'-s)}ds'\ ,\label{fS}
\eqa
 where $L=(-\infty,0]$ and $R$ denotes cuts at higher energies other than the
$2\pi$ elastic cut. To a good approximation $R$ starts at
$4m_K^2$. The parameterization form, Eqs.~(\ref{param}) --
(\ref{fS}), which may be viewed as an improved version of the
Dalitz--Tuan representation, has the advantage that $S^{R_i}$ does
not contribute to the discontinuity of $f$,\cite{piK} which means
the following relation: \bqa\label{IMLR}
 \mathrm{Im}_{L,R}f(s)=-{1\over
{2\rho(s)}}\log|S^{phy}(s)|\ .
\eqa%
The left cut integral in the above equation are estimated  by
replacing $S^{phy}$ on the $r.$$h.$$s.$ of Eq.~(\ref{IMLR}) by
$S^{\chi PT}$ which is available at two--loop order for $\pi\pi$
scatterings~\cite{cpt2}. In Fig.~\ref{ImLf} we plot the function
$\mathrm{Im}_{L}f(s)$ using both tree, 1--loop and 2--loop
$\chi$PT results at moderately low energies from which one may get
an impression on the convergence speed of the chiral expansion.
Evidences exist in the I,J=2,0 channel of $\pi\pi$
scatterings~\cite{XZ00} and I,J=${3\over 2}$,0 channel of $\pi K$
scatterings~\cite{piK} to support such an approximation. The
present method is certainly superior to the method of
Ref.~\cite{XZ00}, since now the constraint of unitarity is
automatically embedded in the formalism. As a bonus of being
unitary, when evaluating the left hand integral using the
following equation, \bqa\label{IMLR'}
 \mathrm{Im}_{L}f(s)=-{1\over
{2\rho(s)}}\log|S^{\chi PT}(s)|\ ,
\eqa%
the once-subtracted dispersion integral automatically converges
owing to the logarithmic form of the integrand. This property
further regulates and suppresses the bad high energy behavior of
the perturbation amplitudes. Nevertheless  it should be pointed
out that though the dispersion integral is formally convergent
when one uses $\chi$PT result to make numerical estimation, the
contribution at high energies is beyond control theoretically due
to the bad high energy behavior of $\chi$PT. Therefore  we
introduce a cutoff parameter when evaluating the left hand
integral in each channel in the  numerical studies, which are
determined by the minimization of $\chi^2$. This is of course,
only a phenomenological approach. Actually, the above
parametrization form can be obtained only by assuming the
analyticity property on the whole cut plane, which is a
consequence of Mandelstam representation. The Mandelstam
representation however goes beyond what is rigorously established
from field theory. Nevertheless the analyticity domain is large
enough for phenomenological applications not very far from the
physical region.~\cite{martin_cheung}

The above discussion briefly reviewed  what is already established
in previous studies. However further improvement is possible. In
$\pi\pi$ scatterings the kinematics is much simpler than in $\pi
K$ scatterings which makes it easy to realize that the subtraction
constant appeared in Eq.~(\ref{param}) can actually be fixed. To
see this we recast Eq.~(\ref{param}) into the following form:
 \be\label{asyms}
\frac{1}{2i\rho(s)}\log(1+2i\rho(s)T^{phy}(s))
=\frac{1}{2i\rho(s)}\sum_i\log(S^{R_i}(s))+f(s)\ .
 \ee
The $l.h.s.$ of the above equation contains the normal branch
points of $S$ as well as the branch points at $z_i$ (which denote
the pole positions of $S$). On the $r.h.s$ of Eq.~(\ref{asyms})
the normal branch points are presented in $f(s)$ whereas the
branch points at $z_i$ are contained in the first term. In the
absence of the long range Coulomb interaction the $\pi\pi$
scattering amplitudes $T^{phy}$ is well behaved when $s\to 0$ and
$4m_\pi^2$. Also it is not difficult to determine the behavior of
$S^{R_i}(s)$ when $s\to 0$ and $4m_\pi^2$. As a consequence, one
obtains, by evaluating the asymptotic behavior of each side of
Eq.~(\ref{asyms}) when $s\to 0$ and $s\to 4m_\pi^2$, the following
relations: \bqa
\label{asy1} &&f(0)=0\ ,\\
\label{asy2}&&a=\sum_ia^{R_i} +f(4m_\pi^2) \eqa
 respectively. In
Eq.~(\ref{asy2}) $a$ is the scattering length of the partial wave
amplitude, $a^{R_i}$ denotes the scattering length contributed by
pole $R_i$. One advantage of using Eq.~(\ref{asy1}) is that it
fixes the subtraction constant appeared in Eq.~(\ref{fS}) and
suggests that the natural subtraction point in Eq.~(\ref{fS}) is
at $s_0=0$, that is, \bqa
 f(s)=\frac{s}{\pi}\int_{-\infty}^{0}\frac{{\rm
 Im}_Lf(s')}{s'(s'-s)}ds'
 +\frac{s}{\pi}\int_{4m_K^2}^{\infty}\frac{{\rm
 Im}_Rf(s')}{s'(s'-s)}ds'\ .\label{fS''}
\eqa
 The validity of Eq.~(\ref{fS''}) can also be demonstrated by
directly evaluating the asymptotic behavior of $f(s)$ when $s\to
0$. Starting from Eq.~(\ref{fS''}) and Eq.~(\ref{IMLR}),
remembering that $T^{phy}(0)$ is a constant, it is not difficult
to demonstrate that the asymptotic behavior of $S^{cut}(s)$ when
$s\to 0$ agrees  with the $l.$$h.$$s.$ of
Eq.~(\ref{param}).\footnote{Up to a sign if the total number of
bound or virtual states in a given channel is odd. } Therefore
Eq.~(\ref{fS''}) is correct.
\begin{center}
\begin{figure}
\mbox{\epsfxsize=6.5cm\epsffile{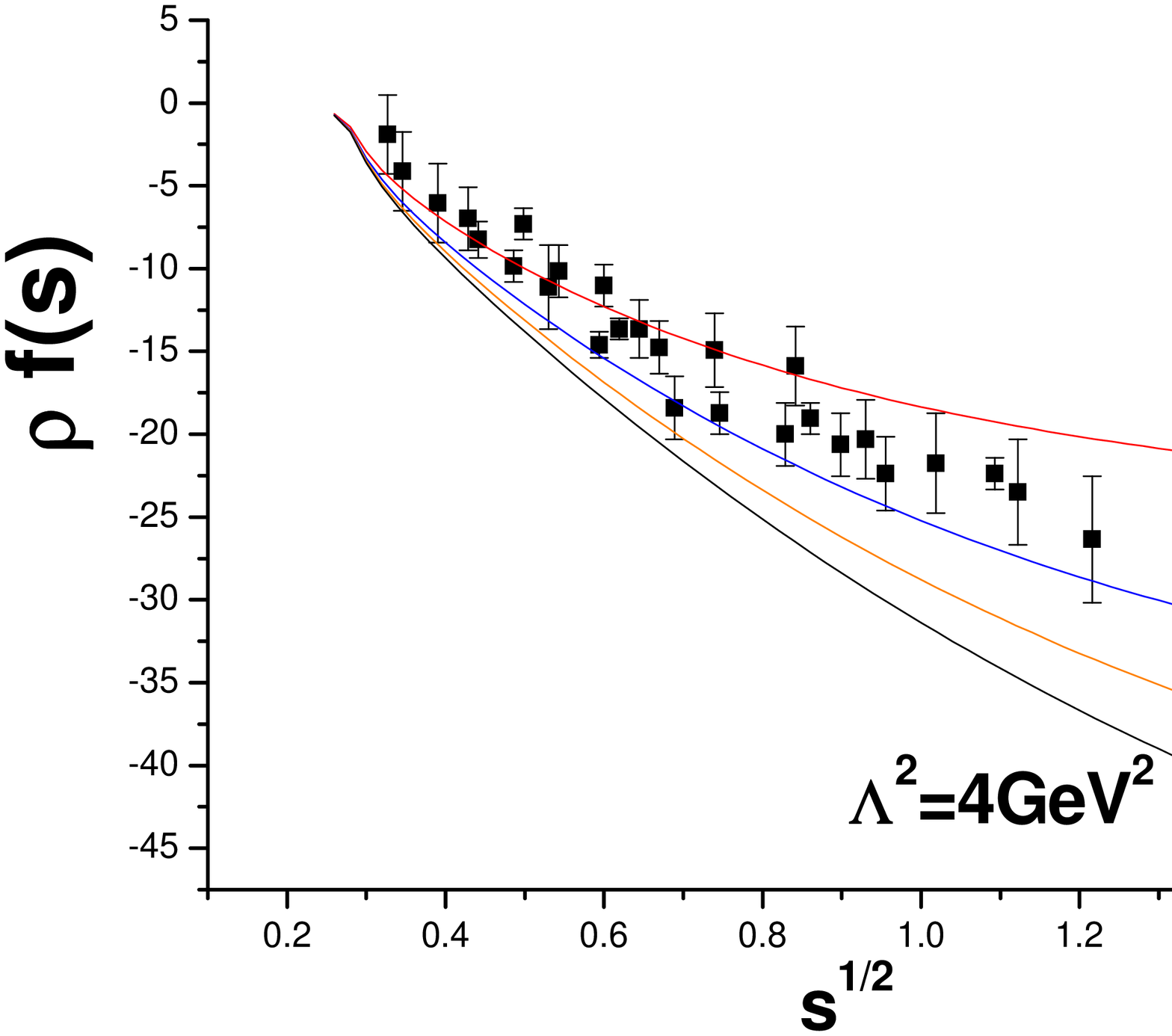}\hspace{1cm}\epsfxsize=6.5cm\epsffile{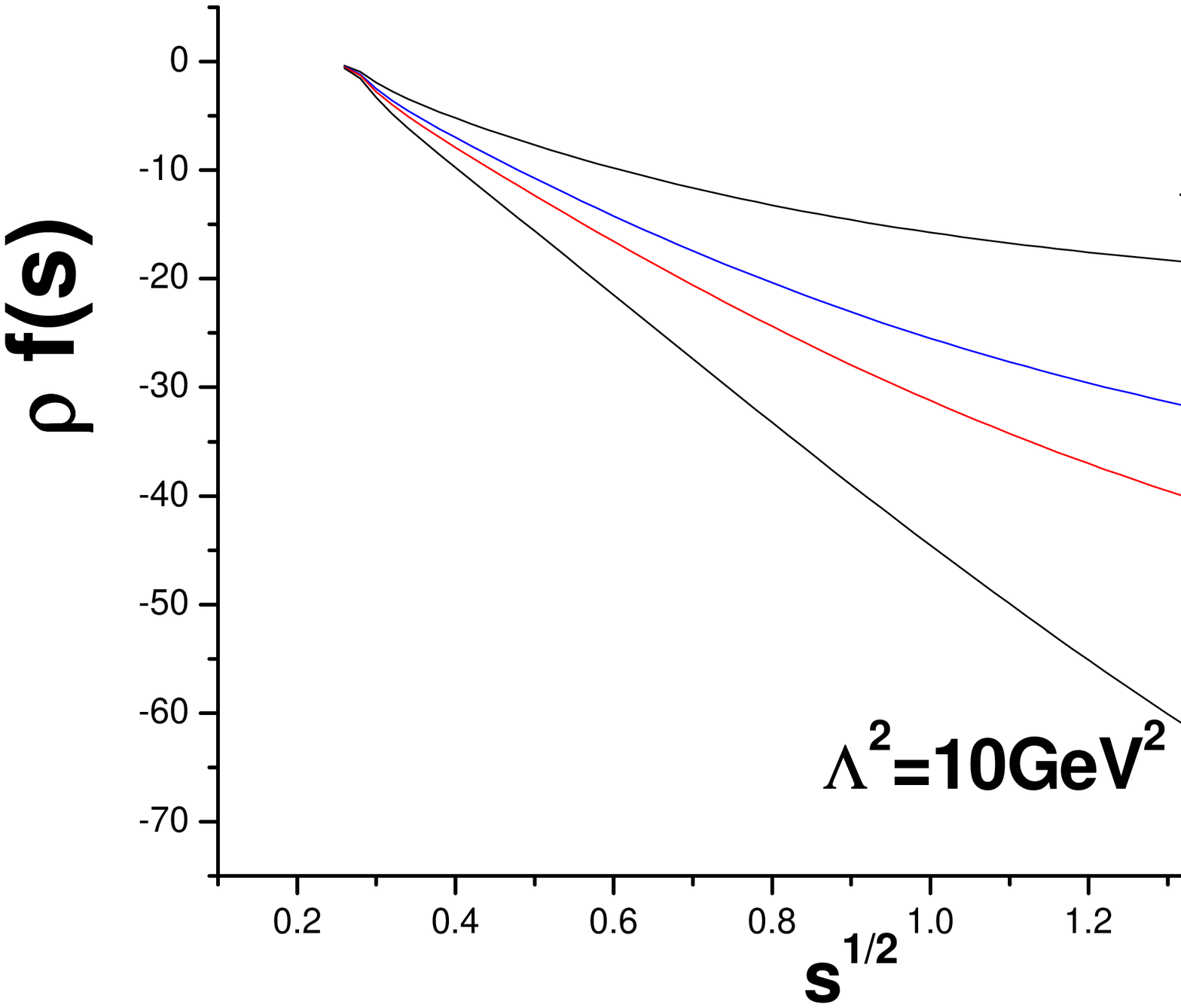}}
\vspace{-2cm} \caption{\label{figf}Different estimates  on the
background contribution to the phase shift. Left: in the I=2, J=0
channel; right: in the I=0, J=0 channel (here the right hand cut
contribution is not included). The experimental data of
$\delta_0^2$ are from Ref.~\cite{Gr74}.}
\end{figure}
\end{center}

The Eq.~(\ref{param}) has been found very useful  in
Ref.~\cite{piK} when studying   the low lying resonance structure
in  $\pi K$ scatterings. Motivated by the discussion given earlier
in this Letter it is worthwhile to re-examine the $\sigma$ pole in
$\pi\pi$ scatterings. For such a propose, it is necessary to
firstly estimate the the background contribution $f$. The left
hand cut integral in Eq.~(\ref{fS''}) is estimated using two loop
chiral perturbation theory results on $\mathrm{Im_L}f$. The
integral interval of the left hand integral in Eq.~(\ref{fS''} )
is taken as $[-\Lambda_L^2,0]$. In Fig.~\ref{figf} we plotted the
cutoff dependence of the function $f$.   For the right hand cut
integral in Eq.~(\ref{fS''}), we neglect its contribution in the
IJ=11 and IJ=20 channels. In the IJ=00 channel we will use a
simple model found in the literature,~\cite{ZB93} which gives a
rather good fit to the inelasticity parameter in the energy region
above $K\bar K$ threshold, to estimate its contribution. The
integration will be truncated at some appropriately chosen cutoff
scale: $\Lambda_R^2\simeq 2$GeV$^2$.

In the following we will  discuss the speciality of IJ=20, 11, 00
channels separately. Then we will include the constraints from
crossing symmetry to make a combined fit to the data of the three
channels below 1GeV and the so called BNR relations.

 \textbf{IJ=20 channel:}

 According to
Eq.~(\ref{asy2}),  if there is no resonance or virtual state in
the IJ=20 channel the scattering length will have a simple
relation: $a^2_0=f(4m_{\pi}^2)$, which is calculable and its value
depends only on $\Lambda_L$ in our scheme. However such a formula
gives a too large magnitude of the scattering length $a^2_0$. For
example, when $\Lambda^2_L$ changes from 0.5GeV$^2$ to
10$^6$GeV$^2$, $f(4m_{\pi}^2)$ varies from $-0.107 $ to $-0.187$,
a value too large comparing with the experimental value
$a^2_0=-0.028\pm 0.012$ ~\cite{EXP79}  and the  lattice
results~\cite{liuc03}, though the lattice method is still not
mature yet. The apparent discrepancy can be solved, recalling that
there actually exists a virtual state pole in the IJ=20
channel.~\cite{ang}
 The pole location can be estimated from the 2 loop perturbative amplitude
 from $\chi$PT to be
 $s_0=0.049m_\pi^2$ on the real axis on the second sheet of
the complex  $s$-plane. According to the definition of pole term
adopted in the present paper, this virtual state pole's
contribution to $a^2_0$ is rather large ($\simeq 0.11$) which can
not be neglected. The existence of the virtual state pole is
unavoidable if chiral perturbation expansion works in the small
$s$ region (actually it is a prediction of current algebra and the
relativistic kinematics~\cite{ang}). In the present approach its
existence is further tested and is found to be fully consistent
with the $S$ matrix unitarization of $\chi$PT. Further, the
existence of such a state raises an interesting question on the
deeper understanding of the physical interpretation of $S$ matrix
poles and the quark hadron duality.

\textbf{IJ=11 channel:}

 The speciality
 in this channel is that the first term of the scattering length expanded in powers of pion
 momentum vanishes:
 \be\label{ab}
\mathrm{Re}T_{IJ}(s)=q^{2J}[a^I_J+b^I_Jq^2+O(q^4)]\ ,\,\,\,\,
(q=\frac{1}{2}\sqrt{s-4m_\pi^2}\,)\ ,
 \ee
  which
 leads to the following constraint:
\begin{equation}\label{sumrule}
f(4m_{\pi}^2)=-\sum_i a^{R_i}\ .
\end{equation}
It is however not convenient to solve this constraint to reduce
one resonance parameter, we therefore make the trick to subtract
$f$ both at $s=0$ and at $s=4m_\pi^2$,
 \bqa
f(s)&=&\frac{f(4m_\pi^2)}{4m_\pi^2}s+\frac{s(s-4m_\pi^2)}{\pi}\int_{L+R}
\frac{\mathrm{Im}f(s')}{s'(s'-4m_\pi^2)(s'-s)}\nonumber\\
&\to& -\sum_i
a^{R_i}\frac{s}{4m_\pi^2}+\frac{s(s-4m_\pi^2)}{\pi}\int_{L+R}\frac{{\rm
 Im}f(s')}{s'(s'-4m_\pi^2)(s'-s)}ds'\ ,
\eqa to meet both the threshold and the $s=0$ behavior of
$f(s)$.~\footnote{We are in debt to I.~Caprini for pointing out
the mistake we made here in the previous edition.}
 There exists apparent disagreement between the phase shift
data~\cite{proto,Estabrooks&Martin}  and theory in this channel,
which results in a
large total $\chi^2_{I=1,J=1}$. 
The contribution to the large $\chi^2_{11}$ comes mainly from low
energy region ($\sqrt{s}\sim0.55$GeV) and also from high energy
region ($\sqrt{s}\sim0.9$GeV), considering the extraordinarily
small error bars of the data.  As a consequence, the $\chi^2_{11}$
is very sensitive to the physical fit parameters as well as
$\Lambda^2_L$ in this channel.
 In addition to the
$\rho(770)$ pole, if we introduce another pole $\rho'$ to simulate
the effects of $\rho(1450)$ and higher resonances it will be
helpful to reduce the large total $\chi^2_{11}$ by roughly 50 and
the $\chi^2_{11}$ is no longer very sensitive to $\Lambda^2_L$.
Also it will improve the fit value of the scattering length and
the effective range parameter defined in Eq.~(\ref{ab}). Therefore
we will add a $\rho'$ pole contribution in the fit.~\footnote{When
introducing the additional pole one has to confine it in the
vicinity of 1450MeV. Otherwise the pole will move down to $M\sim
300$MeV  with a large width to remedy the apparent disagreement
between theoretical curve and data at low energies. This is
unacceptable because the scattering length and the slope parameter
will be totally different from the $\chi$PT results. The
$\rho(1450)$ pole can not be fixed reasonably well in the present
calculation since it suffers from the fact that we do not attempt
to include the data around 1450 MeV, the deficiency of the data at
low energies, and uncertainty of the cut-off integral when
evaluating the background contribution. Therefore to introduce the
$\rho'$ pole only aims at improving the present fit results. }
 The resulting $\chi^2_{11}$  is still very large mainly due to the
apparent disagreement between the theoretical curve and the data
at low energies. Nevertheless the $\rho$ pole position is very
stable against different choices of $\Lambda_L^2$. Also the
$\sigma$ pole position changes very little with or without adding
the $\rho'$ pole (less than 10 MeV) in the combined fit as
described later.

Because the consistency problem between data from
Ref.~\cite{proto,Estabrooks&Martin} and theory as mentioned above,
it is necessary to look for experimental information from other
sources. There exists accurate low energy  data on pion
form-factor in the I=1,J=1 channel from the CMD-2
experiment.~\cite{CMD2}\footnote{We thank Prof. Leutwyler for
bringing Ref.~\cite{CMD2} to our attention.} However since what we
need is the phase shift data, the data provided by CMD-2
experimental group are not directly usable here. The way we
extract the `experimental' data on $\delta_1^1$ is the following:
we use Eq.~(8) in Ref.~\cite{CMD2} to refit the form-factor data
to determine parameters, then we turn off the $\rho$--$\omega$
mixing parameter $\delta$ and calculate the phase shift
$\delta_1^1$ using again Eq.~(8) of Ref.~\cite{CMD2}. In this way
we can manipulate the `experimental' data on $\delta_1^1$. It is
arbitrary to choose the center of mass energy of each data point,
nevertheless we fix the center of mass energies of the `data' set
the same as the original data for form-factors provided by
Ref.~\cite{CMD2}. The error bars of the phase shift `data' are
calculated using the error matrix method. In this way we obtain 43
`data' points for $\delta_1^1$.

If using our formula to fit the manipulated $\delta_1^1$ data
described above we get a very small total $\chi^2_{11}$, which
suggests that there may exist strong correlations between the
error bars of the manipulated data and/or that our formula and the
Eq.~(8) of Ref.~\cite{CMD2} are highly consistent. Though there
appears drastic difference on total $\chi^2$ between the data of
Ref.~\cite{proto} and the data manipulated from Ref.~\cite{CMD2},
it will be shown later that the difference has only small
influence on the determination of the $\sigma$ pole position.

 \textbf{IJ=00 channel:}

Similar to the argument given in Ref.~\cite{XZ00}, the cut
contribution calculated from Eq.~(\ref{fS''}) is negative, which
clearly  demonstrates that the $\sigma $ resonance must exist to
reproduce the observed experimental phase shift in the I=0, J=0
channel, which is positive and rising at low energies. The
determination of the pole position of the $\sigma $ resonance is
 straightforward.
 In the I=0, J=0 channel we begin by introducing
 two poles below 1GeV region: one is the $\sigma $
resonance under investigation and the another one corresponds to
$f_0(980)$.
 With their masses and
widths as fit parameters, MINUIT displays an obvious broad
resonance and a narrow resonance. However, cautious readers may
still ask the question whether the strong enhancement of
$\delta_0^0$ below 1GeV is owing to the sum of  residue effects of
many resonances located far away from the physical region other
than from a single $\sigma$ pole. To explore such a possibility we
introduce a third pole without any prejudice on where it should
locate. We will test whether the $\sigma$ pole location will be
severely  distorted by the introduction of this additional pole.
The answer is no. The $\sigma$ pole  moves only slightly with or
without  the third pole included (roughly less than 20 MeV). On
the other side the location of the third pole is found to be very
flexible when varying $\Lambda_L$, but always stays far away on
the left. Roughly we find $M_{\mathrm{3}}$ ranges from a few tens
of MeV to a few hundreds of MeV whereas $\Gamma_{\mathrm{3}}$
ranges from 1 - 3GeV. Such a pole would better be described as
pure background and its contribution to the phase shift remains
small.
\begin{figure}%
\begin{center}%
\mbox{\epsfxsize=10cm\epsffile{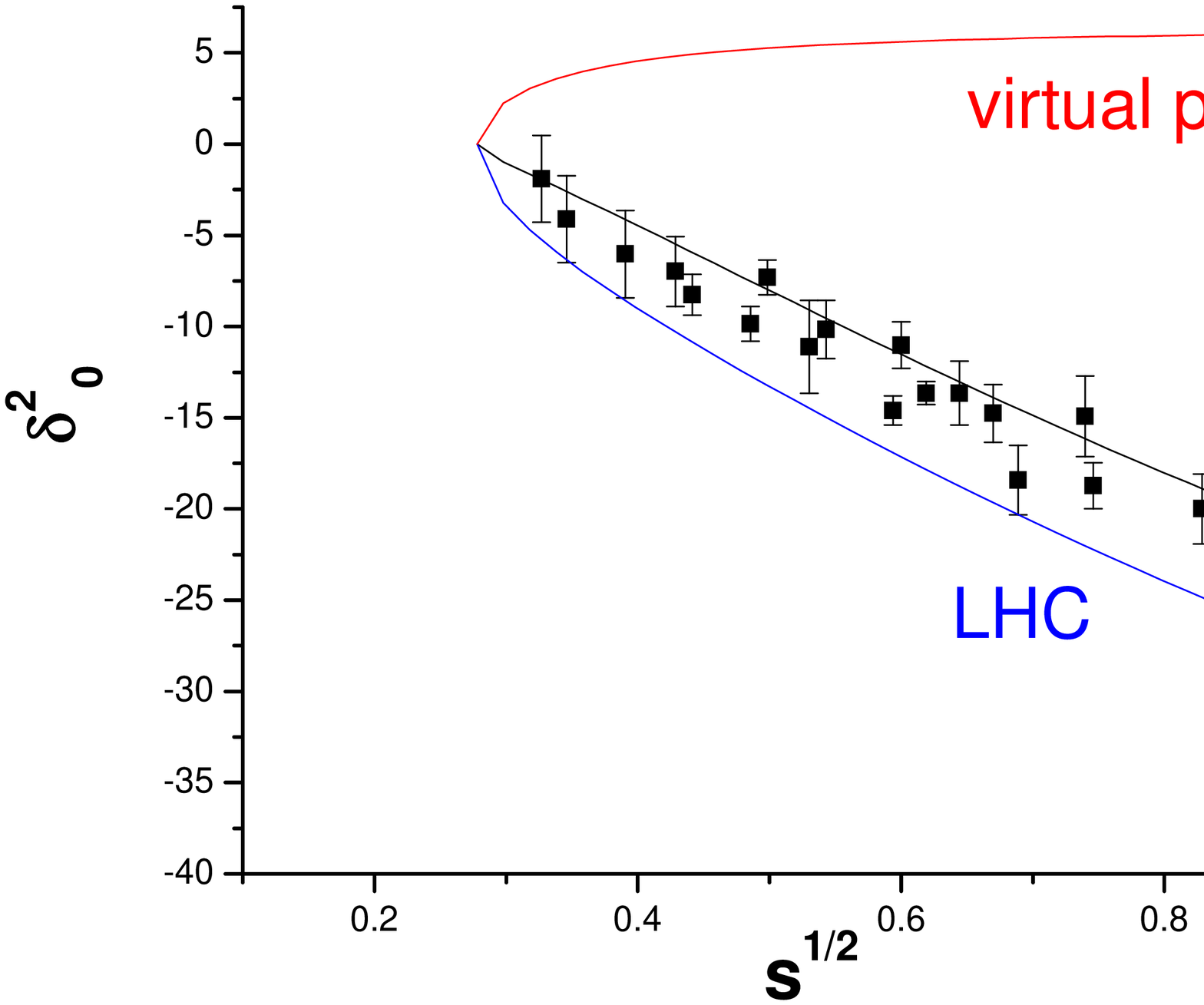}}
\mbox{\epsfxsize=10cm\epsffile{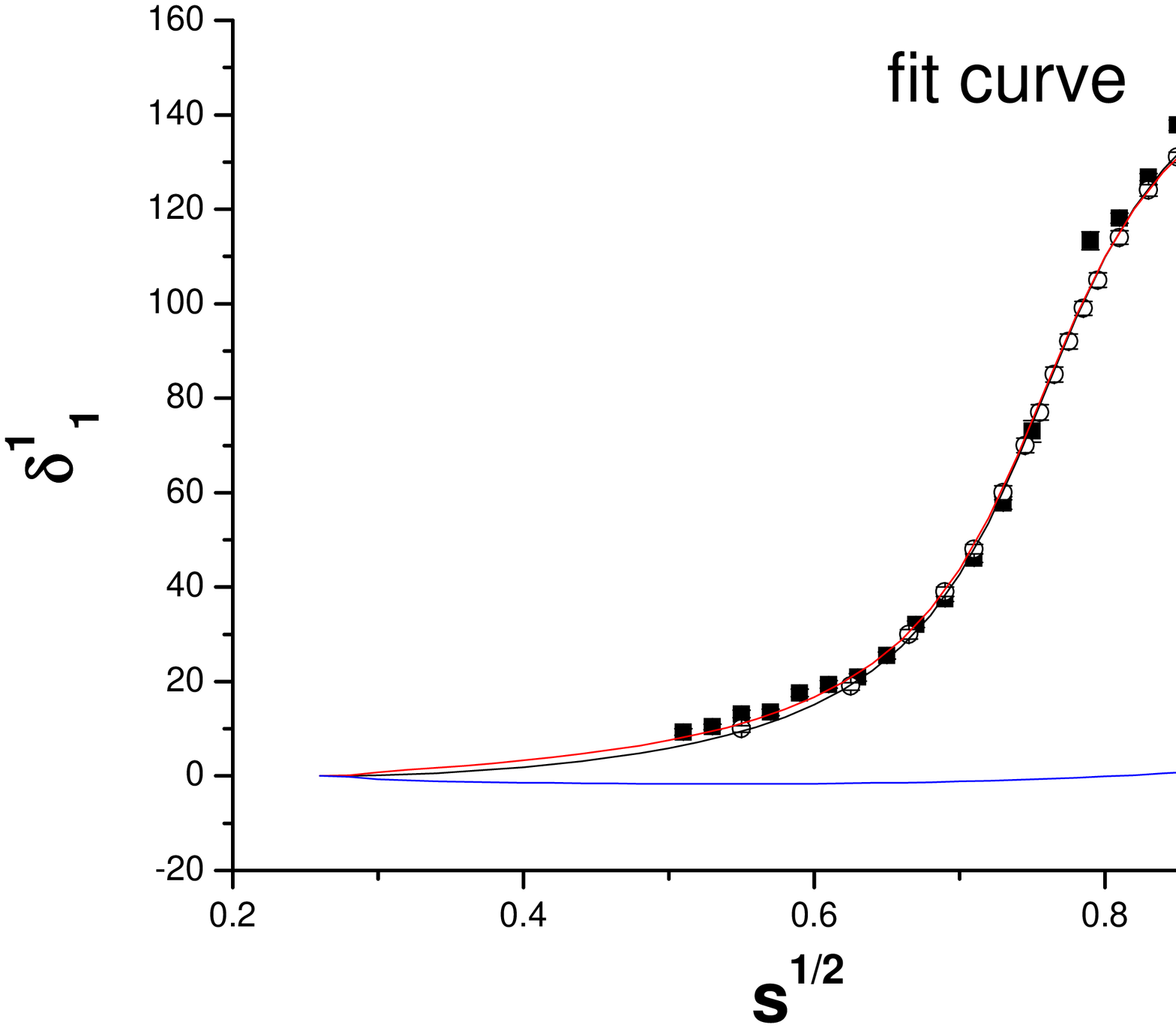}}
\vspace{-1cm}\mbox{\epsfxsize=10cm\epsffile{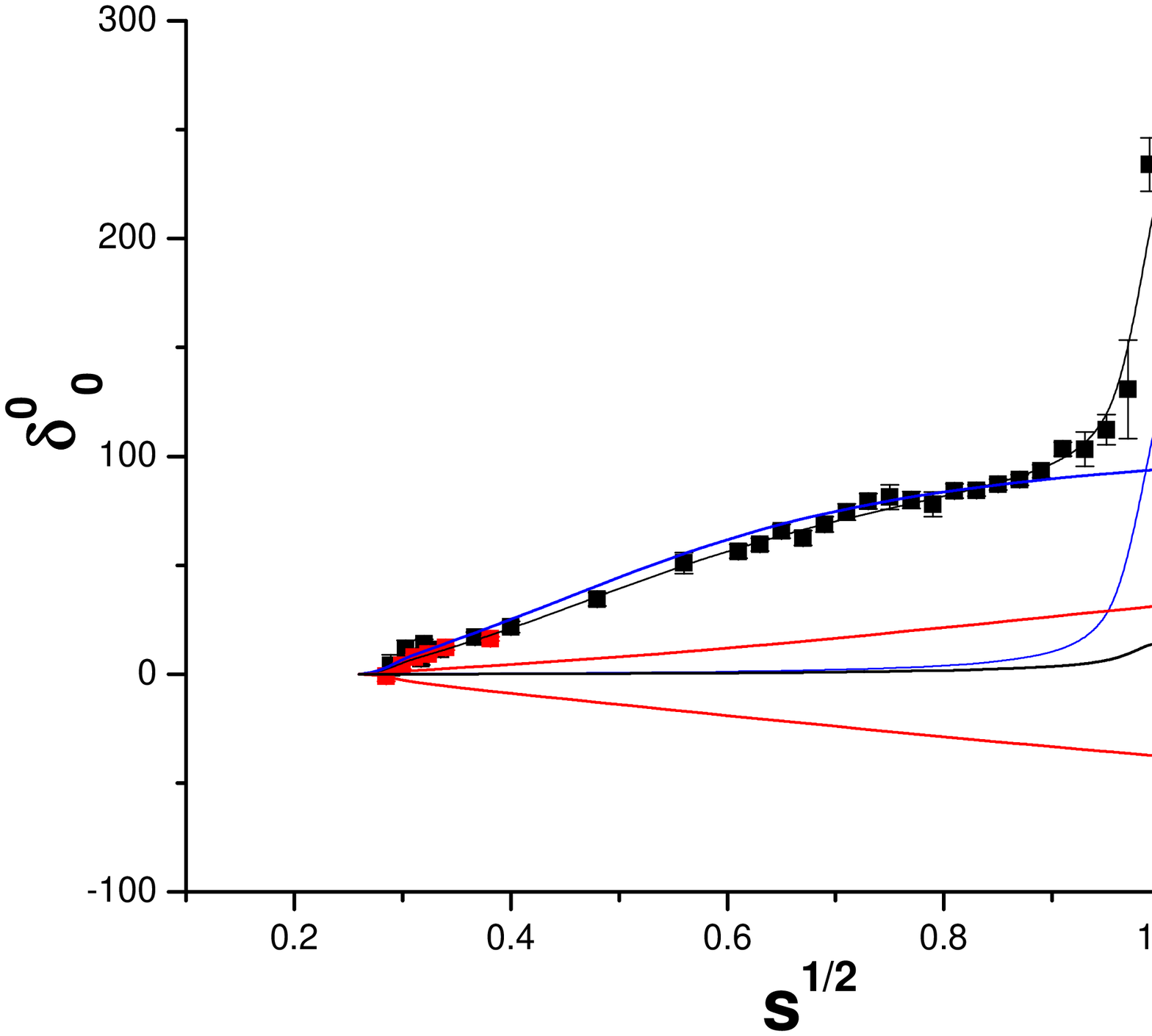}}
\vspace{-1cm}\caption{\label{figphase}Fit results on phase shifts
from  Eq.~(\ref{result}). Data in the IJ=11 channel are: open
circle from Ref.~\cite{proto}, full square from
Ref.~\cite{Estabrooks&Martin}. Data in the other channels are
described in the text.}
\end{center}%
\end{figure}%

\textbf{The combined fit with constraints from crossing symmetry}

A major shortcoming of the unitarized partial wave amplitudes as
described above is that crossing symmetry is not automatically
built in, despite of its apparent advantages. On the other side,
necessary conditions involving only low partial waves can be
derived from crossing symmetry. In $\pi\pi$ scatterings, there
exists crossing relations between partial wave amplitudes -- BNR
relations. There exists five such relations if only $s$-wave and
$p$-wave are considered, i.e.,
 \bqa\label{BNR1}
\mathrm{I}:\,\,\,&&\int_0^4(s-4)(3s-4)[{t_0^0}(s)+2{t_0^2}(s)]ds=0\
,\nonumber\\
\mathrm{II}:\,\,\,&&\int^4_0(s-4)R_0^0[2t^0_0(s)-5t^2_0(s)]ds=0\
,\nonumber\\
\mathrm{III}:\,\,\,&&\int^4_0(s-4)R_1^0[2t^0_0(s)-5t^2_0(s)]ds-9\int^4_0(s-4)^2R^1_0t^1_1(s)=0\
,\nonumber\\
\mathrm{IV}:\,\,\,&&\int^4_0(s-4)R_2^0[2t^0_0(s)-5t^2_0(s)]ds+6\int^4_0(s-4)^2R^1_1t^1_1(s)=0\
,\nonumber\\
\mathrm{V}:\,\,\,&&\int^4_0(s-4)R_3^0[2t^0_0(s)-5t^2_0(s)]ds-15\int^4_0(s-4)^2R^1_2t^1_1(s)=0\
, \eqa where $R_i^j$ are,
 \bqa\label{BNR2}
&&R_0^0=1\ ,\,\,\,R_0^1=1\ ,\nonumber\\
&&R_1^0=3s-4\ ,\,\,\,R_1^1=5s-4\ ,\nonumber\\
&&R_2^0=10s^2-32s+16\ ,\,\,\,R^1_2=21s^2-48s+16\ ,\nonumber\\
&&R^0_3=35s^3-180s^2+240s-64 \ .\eqa Dimensional parameters in
Eqs.~(\ref{BNR1}-\ref{BNR2}) are expressed in unit of $m_\pi^2$
for simplicity. All these BNR relations can be rewritten in a
compact form:
$$\int^4_0ds[P^i_{00}(s)T^0_0(s)+P^i_{11}(s)T^1_1(s)+P^i_{20}(s)T^2_0(s)]=0\ ,$$
where $i$ ranges from I to V and $P^i_{IJ}(s)$ are polynomials. We
will make a combined fit on the IJ=20,11,00 channels together with
the 5 BNR relations using the penalty function method.~\cite{zhu}
We define the penalty function as follows:
 \bqa\label{BNRERR}
\chi^2_{BNR}&\equiv&\frac{1}{\epsilon^2}\sum_{i=I}^V(Err^{(i)})^2\nonumber\\
&\equiv&\frac{1}{\epsilon^2}\sum_{i=I}^V
\frac{|\int^4_0ds[P^i_{00}(s)T^0_0(s)+P^i_{11}(s)T^1_1(s)+P^i_{20}(s)T^2_0(s)]|^2}
{[\int^4_0ds(|P^i_{00}(s)T^0_0(s)|+|P^i_{11}(s)T^1_1(s)|+|P^i_{20}(s)T^2_0(s)|)]^2}\
,\nonumber\\\eqa and define the total $\chi^2$ as, \be
 \chi^2_{tot}=\chi^2_{20}+\chi^2_{11}+\chi^2_{00}+\chi^2_{BNR}\ .
\ee
 Altogether we have 10 physical parameters corresponding to 5 poles
 (3 in the IJ=00 channel and 2 in the IJ=11 channel). In addition
 we have three cutoff parameters when evaluating the background contributions
 in three channels.  The penalty factor,
$\epsilon^{-2}$, varies between $10^{-4}$ to $10^{-5}$. The rule
to choose such a range is that we require that the violation of
each BNR relation (corresponding to the value of $Err^{(i)}$
defined in Eq.~(\ref{BNRERR})) is at the level of one percent.

In the following combined fit we will test different combination
of data. Firstly  we use the data of IJ=11 channel from
Ref.~\cite{proto,Estabrooks&Martin}. Below 1 GeV there are 23 data
points in the IJ=20 channel~\cite{Gr74}, 39 data points in the
IJ=11 channel~\cite{proto,Estabrooks&Martin} and 36 data
points~\cite{Gr74,oldKe4,newKe4} in the IJ=00 channel. In the fit
it is found that the total $\chi^2$ is  sensitive to
$(\Lambda^2_L)_{IJ=20}$ and the best choice corresponds to
$(\Lambda^2_L)_{IJ=20}\simeq 3.4$GeV$^2$. On the other side the
total $\chi^2$ is no longer sensitive to $(\Lambda^2_L)_{IJ=11}$
when the $\rho'$ pole is included. Also the total $\chi^2$ and all
major physical outputs are insensitive to $(\Lambda^2_L)_{IJ=00}$.
Part of the reason behind such a phenomenon is because in the
IJ=00 channel the third pole's location is sensitive to the choice
of the cutoff parameter and it absorbs the cutoff dependence of
the background integral. It is found that the $\sigma$ pole
location is almost inert against the variation of the cutoff
parameter. Though the third pole's position is very flexible, it
always wanders on the far left side of the complex $s$ plane.
Taking the penalty factor $\epsilon^{-2}=10^{-4}$ is enough to
constrain the violation of BNR relations at the level of 1
percent, Further decreasing  it down to $10^{-5}$ does not help
much in improving those BNR relations but it causes further
problems on the stability of the global fit, though the center
value of the $\sigma$ pole position changes very little. In below
we list the fit results corresponding to (almost) the minimal
$\chi^2$ we can find:
 \bqa\label{result}
&&\chi^2_{tot}=\chi^2_{00}+\chi^2_{11}+\chi^2_{20}+\chi^2_{BNR}=29.7+214.9+41.6+4.4\ ;\nonumber\\
&&M_\rho=757.0\pm 0.4MeV\ ,\,\,\, \Gamma_\rho=152.2\pm 0.6MeV\ ,\nonumber\\
&&M_{f^0}=984.5\pm 2.3MeV\ ,\,\,\, \Gamma_{f^0}=34.4\pm 6.8MeV\ ,\nonumber\\
&&M_\sigma=457\pm 15MeV\ ,\,\,\, \Gamma_\sigma=551\pm 28MeV\ .
 \eqa
 The third pole location in the IJ=00 channel is $M_{3}=819$MeV,
 $\Gamma_{3}=1846$MeV with large error bars.  The fit results are plotted in
 Fig.~\ref{figphase}.
 The obtained
scattering lengths  and the effective range parameters are listed
in table~\ref{tab2} where we also list several results found in
the literature for comparison. From table~\ref{tab2} we find that
our results are in general in good agreement with the $\chi$PT
results~\cite{BCT98} and especially the Roy equation
analysis~\cite{CGL01}. However, since  we used the flawed low
energy data in the IJ=11 channel,\footnote{if we do not include
the so called $\rho'$ pole the $b_1^1$ parameter does not agree
well with the Roy equation analysis. } we will also test in the
following the manipulated data of $\delta_{1}^1$ from CMD-2.
Nevertheless the problem of the data in the IJ=11 channel does not
seem to propagate into IJ=00 and 20 channels in our approach,
since the constraints from the five BNR relations on the IJ=11
channel are rather weak because the magnitude of $T_{11}$ is
suppressed in the region governed by the BNR relations.
 {\footnotesize \begin{table}[bt]
\centering\vspace{0.1cm}
\begin{tabular}{|c|c|c|c|c|c|}
\hline $\ $&Our\ results&$\chi$PT~\cite{BCT98}&Roy
Eqs.~\cite{CGL01}&Exp.~\cite{EXP79}&Unit
\\ \hline
$a^0_0$&$0.211\pm 0.011$&$0.220\pm 0.005$&$0.220\pm
0.005$&$0.26\pm 0.05$&
\\
$b^0_0$&$0.264\pm 0.015$&$0.280\pm 0.011$&$0.276\pm
0.006$&$0.25\pm 0.03$&$ m_\pi^{-2}$
\\ \hline
$a^2_0$&$-0.440\pm 0.011$&$-0.423\pm 0.010$&$-0.444\pm
0.010$&$-0.28\pm 0.12$&$10^{-1} $
\\
$b^2_0$&$-0.785\pm 0.010$&$-0.762\pm 0.021$&$-0.803\pm
0.012$&$-0.82\pm 0.08$&$10^{-1} m_\pi^{-2}$
\\ \hline
$a^1_1$&$0.367\pm 0.003$&$0.380\pm 0.021$&$0.379\pm
0.005$&$0.38\pm 0.02$&$10^{-1} m_\pi^{-2}$\\
$b^1_1$&$0.563\pm 0.003$&$0.58\pm 0.12$&$0.567\pm 0.013$&$\
$&$10^{-2} m_\pi^{-4}$
\\ \hline
\end{tabular}
\caption{\label{tab2}Threshold parameters comparing with other
works. Phase shift data in the IJ=11 channel are from
Ref.~\cite{proto,Estabrooks&Martin}. More accurate experimental
results of $a_0^0$ and $a^2_0$ of most recent $K_{e4}$ experiment
 by E865 Collaboration can be found in reference \cite{newKe4}
: $a^0_0=0.216\pm0.013({\rm stat. })\pm0.004({\rm
 syst. })\pm0.002({\rm theor.})$, $a^2_0=-0.0454\pm0.0031({\rm stat. })\pm0.0010({\rm
 syst. })\pm0.0008({\rm theor.})$. See
also the first reference of \cite{PY03} for a different analysis
of the same data.}
\end{table}}

In the second approach we  use the manipulated data in the IJ=11
channel from Ref.~\cite{CMD2}. The solution is the following:
 \bqa\label{result'}
&&\chi^2_{tot}=\chi^2_{00}+\chi^2_{11}+\chi^2_{20}+\chi^2_{BNR}=29.8+0.02+41.6+5.9\ ;\nonumber\\
&&M_\rho=763.0\pm 0.2MeV\ ,\,\,\, \Gamma_\rho=139.0\pm 0.5MeV\ ,\nonumber\\
&&M_{f^0}=984.4\pm 2.7MeV\ ,\,\,\, \Gamma_{f^0}=34.6\pm 7.9MeV\ ,\nonumber\\
&&M_\sigma=459\pm 3MeV\ ,\,\,\, \Gamma_\sigma=551\pm 23MeV\ .
 \eqa
 The obtained
scattering lengths  and the effective range parameters in the
IJ=00 and 20 channels are found to be in satisfactory agreement
with those listed in table~\ref{tab2}. However  in the I,J=1,1
channel we find that  $a_1^1=3.54\times 10^{-2}$, $b_1^1=4.27
\times10^{-3}$, which does not seem to be compatible with the
results obtained from the Wanders sum rule given in
Ref.~\cite{ACGL}. Nevertheless we find, by comparing
Eq.~(\ref{result}) with Eq.~(\ref{result'}), that the $\sigma$
pole location changes very little.

 Finally we briefly comment on the use of another group of  IJ=20
 data by Hoogland et al..\cite{Hoogland} The quality of solution B
 of Ref.~\cite{Hoogland} is not good as it gives a very large $\chi^2$. For
 solution A, 
 in the combined
 fit the central value of $a^2_0$ is $-4.09\times10^{-2}$. The other outputs are compatible
  to previous results
 and the fit results in a  satisfactory
 $\chi^2_{BNR}$ and $\chi^2_{20}$. On the other side, the mass and width
 of the $\sigma$ resonance,  will both increase roughly by 30 MeV.

A careful analysis  was performed to estimate the influence of
various theoretical uncertainties to the determination of the
$\sigma$ pole, including the penalty factor $\epsilon$, various
cutoff parameters, different treatment on IJ=11 (and also IJ=20)
channel data, and whether or not including the third pole. It is
found that
 \be\label{rrr1}
M_\sigma=470\pm 50MeV\ ,\,\,\, \Gamma_\sigma=570\pm 50MeV\ ,
 \ee
 where the error bars are rather conservative. The
 results are found in very good agreement with the results given in
 Ref.~\cite{CGL01}. Different from Ref.~\cite{CGL01}, the error
 bars here are mainly caused by the use of different data, rather than
 by analytic continuation.

To conclude, using the  unitarization approach developed by our
group and combining  the constraints from crossing symmetry, it is
found that a fit to the low energy data gives  reasonable results,
comparing with Roy equation analyses.~\cite{CGL01,PY03} The
existence of the virtual state pole in the IJ=20 channel is
reconfirmed. It is also found that crossing symmetry is crucial in
correctly determining the $\sigma$ pole location. Actually using
the present formulae, without the constraints from BNR relations
the individual fit in the IJ=00 channel would give a pole located
at $M_\sigma=542$MeV, $\Gamma_\sigma=546$MeV. It is also found
that the BNR relations provide a stronger correlations between the
IJ=00 and 20 channels and the 11 channel  amplitude plays a
relatively minor role in the BNR relations being used.  The fit
results end up in general in good agreement with the results of
Ref.~\cite{CGL01}. The estimated value of $\sigma$ pole position
as quoted by the Ref.~\cite{PDG02} is certainly unsatisfactory. It
is desirable that newer estimates on the  pole location begin to
converge.

{\bf Acknowledgements:} We would like to thank Profs. K.~L.~He,
D.~H.~Zhang for helpful discussions on data fittings. Also we are
grateful  to Prof. Leutwyler  for valuable comments and critical
remarks. This work is supported in part by China National Natural
Science Foundation under grant number 10491306.

\end{document}